\documentclass[pra,aps,twocolumn,showpacs,amsmath,amssymb]{revtex4}

\usepackage{graphicx}

\begin{document} 
\title{Storing quantum information in a solid using dark state polaritons}
\author{Mattias Johnsson and Klaus M{\o}lmer} 
\affiliation{QUANTOP - Danish National Research Foundation Center for
Quantum Optics, Department of Physics and Astronomy, University of \AA rhus, 
DK 8000 \AA rhus C., Denmark} 
\date{\today} 
 
\begin{abstract} 
The possibility of using a solid medium to store
few-photon laser pulses as coupled excitations
between light and matter is investigated. The role of
inhomogeneous broadening and nonadiabaticity are 
considered, and conditions governing the feasibility 
of the scheme are derived. The merits of a number
of classes of solid are examined.
\end{abstract} 
 
 
\maketitle

\section{Introduction}

In the last few years, many exciting effects in nonlinear
optics have been made possible by using 
electromagnetically induced transparency (EIT). This allows
a near-resonant probe field to experience extreme 
nonlinearities, while simultaneously using a 
second coupling field to cancel the associated
absorption \cite{harris1997, arimondo1996}.
Applications include nonlinear optics at low light levels
\cite{harrisET1998, harrisET1999, lukinET2000, johnssonET2002a}, full 
frequency conversion in distances so short that phase
matching is not relevant \cite{jainET1996, korsunskyET2002,
merriamET2002}, and quantum information storage
\cite{kocharovskayaET2001, fleischhauerET2002}.

Another effect EIT allows is an 
extreme reduction of the group velocity of a light pulse
\cite{arimondo1996,harrisET1999,marangos1998}.
This arises from the very large dispersion
experienced by a probe field that is close to
resonance. Slow light has been theoretically analyzed
and experimentally demonstrated in gaseous media
\cite{kashET1999}, BECs \cite{hauET1999} and in
solids \cite{kuznetsovaET2002,bigelowET2003}.
Group velocities of a few tens of meters per second
have been achieved.

If the coupling field is allowed to become time-dependent,
then the possibility of completely stopping and trapping the 
probe pulse arises \cite{fleischhauerET2000, phillipsET2001,
fleischhauerET2002, liuET2001,turukinET2002}.
To do this one adiabatically reduces
the coupling field to zero while the probe pulse
is within the EIT medium. This results in the transfer
of the probe pulse into a collective spin coherence
between the atoms in the medium. As adiabatic
passage techniques \cite{bergmannET1998} are used, the collective 
atomic state storing the excitation is a ``dark state'',
and contains no component of an upper
level state which can decay. The lifetime of
the dark state is thus governed by the ground
state coherence dephasing rate, which can 
be as low as a few tens of Hertz. In the quantum
picture these coupled electromagnetic and atomic
excitations are best described as a single
dressed state excitation which has been termed
a dark state polariton by Fleischhauer and Lukin
\cite{fleischhauerET2002}.
If the coupling field is subsequently adiabatically
increased back to its original value, 
the spin coherence is transferred back into
the electromagnetic field. The probe pulse is
thus reformed and can propagate further.

As this scheme preserves the quantum state of
the pulse, it allows the possibility of using
such a method for quantum information storage
and processing \cite{fleischhauerET2002}. In addition,
because the quantum state of the input pulse
is mapped onto a many-atom collective excitation,
it does not suffer from fundamental problems preventing
the efficient coupling of a field to a single atom \cite{enkkimble},
and the scheme is robust and immune to
many perturbing effects that can affect storage
schemes utilizing single atoms in the
context of cavity QED \cite{cirzol,grangier}.

Although a growing body of literature is beginning
to consider storing classical light pulses in a solid
\cite{kuznetsovaET2002,turukinET2002},
a theoretical analysis of storing quantum information
in a solid using the dark state polariton
formalism has not been carried out. Such
a scheme would be well worth considering, as
solids have a number of advantages over gases.
They are easy to prepare and store, stored
information does not degrade due to atomic
diffusion, and, above all much higher
densities of interacting atoms can be
achieved. For example, a common class of solids used
within a quantum information context is
rare-earth doped crystals, where the concentration
of dopants can easily exceed the density of atoms
in a gas by eight orders of magnitude.
Outside this class of materials, nitrogen-vacancy
centers in diamond have also been considered
\cite{hemmerET2001, shahriarET2002}. These have
the advantages of a strong oscillator strength
and relatively long spin coherences.
It is also conceivable that one could use doped
glasses instead of crystals, although extreme
inhomogeneous broadening must then be overcome.

The purpose of this paper is to extend the analysis of 
Fleischhauer and Lukin by considering the behavior
of dark state polaritons in a solid,
rather than gaseous, medium. We determine
whether quantum information storage is
still possible, given the large inhomogeneous
broadening that is present in solids,
derive conditions that must be met for 
successful storage, and finally discuss in which
classes of systems the conditions can be
met.

\section{Basic model}

\begin{figure}[ht]
\begin{center}
  \includegraphics[width=7cm]{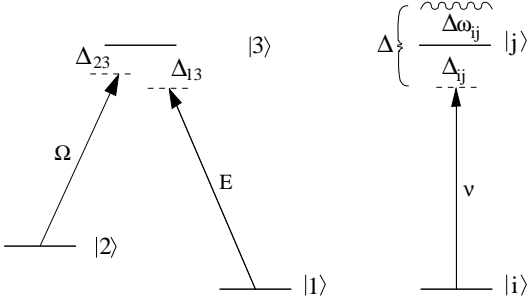}
\caption{Left: three level scheme. $\hat{E}$ is 
a weak quantum field while $\Omega$ is a classical field.
Right: Detuning scheme for a particular atom. The wavy
line represents the inhomogeneous line center for 
the atomic ensemble, thus $\Delta\omega_{ij}$ 
represents the detuning between the upper level
of a particular atom and the inhomogeneous line center.}
\label{fig3level}
\end{center}
\end{figure}

We consider a standard three-level lambda system, 
as shown in Fig. \ref{fig3level}. $\hat{E}$ is
a weak quantum field, while $\Omega$ is the 
Rabi frequency associated with a strong
classical coupling field. We assume that both
fields propagate parallel to the $z$ direction, 
reducing the system to a 1D problem. 

Using a similar approach to \cite{johnssonET2003},
one can show that in the continuum limit the 
interaction Hamiltonian for this system is
given by
\begin{eqnarray}
\hat{H}_{\mathrm{int}} &=& -\frac{\hbar N}{L} 
	\int {\mathrm{d}}z \, [ g\sigma_{21}(z,t)
   \hat{E}(z,t) \nonumber \\
&&	\hspace{1.5cm}+ \Omega(z,t) \sigma_{23}(z,t) 
		+ {\mathrm{H.c.}} ] 
\end{eqnarray}
where all quantities are taken to be slowly varying,
both in time and space, i.e. we have transformed to
a rotating frame. The coupling constant is given by
$g=d_{13} \sqrt{\nu / 2\hbar \epsilon_0 V}$,
where $V$ is the interaction volume and $d_{13}$
is the electronic dipole moment between states 
$|1\rangle$ and $|3\rangle$. $N$ is the total number
of dopant atoms in the interaction volume.

Within the slowly varying amplitude and phase
approximation, the equation of motion for
the quantum field $\hat{E}$ is given by
\begin{equation}
\left( \frac{\partial}{\partial t} + c\frac{\partial}
	{\partial z}\right) \hat{E}(z,t)
	= igN\sigma_{13}(z,t) \label{eqEprop}
\end{equation}

Using variables that are slowly varying in space and time
one finds the atomic equations of motion for a 
single atom to be
\begin{eqnarray}
\dot{\sigma}_{11} &=& -\gamma_{1}\sigma_{11} + ig(\hat{E}^{\dagger}\, 
	\sigma_{13} - \hat{E} \, \sigma_{31}) + F_1 \\
\dot{\sigma}_{22} &=& -\gamma_{2}\sigma_{22} 
	+ i(\Omega^* \,\sigma_{23} - \Omega \,\sigma_{32}) + F_2\\
\dot{\sigma}_{33} &=& -\gamma_{3}\sigma_{33} + i(g \hat{E} \sigma_{31} 
	- g\hat{E}^{\dagger} \,\sigma_{13} \nonumber \\
&& \hspace{1cm} + \Omega \,\sigma_{32} - \Omega^* \,\sigma_{23}) + F_3\\
\dot{\sigma}_{13} &=& \Gamma'_{13} \sigma_{13}
%
+ i(g\hat{E}(\sigma_{11} -
	 \sigma_{33}) + \Omega \,\sigma_{12}) + F_{13}\\
\dot{\sigma}_{23} &=&  \Gamma'_{23} \sigma_{23}
	 + i(g\sigma_{21} \hat{E} + \Omega(\sigma_{22}
	  - \sigma_{33})) + F_{23} \\
\dot{\sigma}_{12} &=& \Gamma'_{12} \sigma_{12}
	+ i(\Omega^* \sigma_{13} - g\hat{E} \,\sigma_{32}) + F_{12}.
\end{eqnarray}
The atomic operators are defined by
$\sigma_{ij} = |i\rangle\langle j|$, $\gamma_i$
represents the population decay from state $|i\rangle$,
and the detunings are defined by
\begin{eqnarray}
\Gamma'_{13} &=& -i\Delta_{13}- \gamma_{13} = -i(\Delta + \Delta\omega_{13}) - \gamma_{13}  \label{eqdeltaomega13} \\
\Gamma'_{23} &=& -i\Delta_{23}- \gamma_{23} = -i(\Delta_0 + \Delta\omega_{23})  - \gamma_{23}  \label{eqdeltaomega23} \\
\Gamma'_{12} &=& -i\Delta_{12}- \gamma_{12} = -i(\Delta-\Delta_0 + \Delta\omega_{12})  - \gamma_{12} \label{eqdeltaomega12}.
\end{eqnarray}
$\gamma_{ij}$ represents the coherence decay between
states $|i\rangle$ and $|j\rangle$; $\Delta\omega_{ij}$
is the detuning of the inhomogeneously broadened
line center from an isolated atom line center.
$\Delta$ and $\Delta_0$ represent the detuning
of the laser from the inhomogeneous line
center for the $|1\rangle$ - $|3\rangle$ transition 
and the $|2\rangle$ - $|3\rangle$ transition respectively.

The $F_{ij}$ are $\delta$-correlated Langevin noise
operators, and as such can be neglected in the
adiabatic limit. We intend to remain close to
this regime. It can also be noted that the 
magnitude of the noise correlations is related
to the atomic decay via the 
fluctuation-dissipation theorem. As the essence
of the transfer process involves utilizing dark
states, ideally the upper state $|3\rangle$ is never
populated. As there is no dissipation, correlations
involving the noise operators vanish. Although as
one moves away from the purely adiabatic regime
an admixture of the bright state with a $|3\rangle$
component becomes excited, this component remains
small, as will be seen in Section 
\ref{subsecNonadiabaticCorrections}. 
Consequently we omit the writing of the noise
operators in the analysis that follows.

We solve the atomic equations of motion perturbatively,
using the expansion parameter $\epsilon =g\hat{E}/\Omega \ll 1$.
Further, we assume that initially all the atoms are
in state $|1\rangle$, so that the zeroth order solutions
for the atomic variables are $\sigma^0_{ij}=1$ if $i=j=1$,
and $\sigma^0_{ij} = 0$ otherwise.

To first order in $\epsilon$ we find
\begin{eqnarray}
\sigma_{12}  &=& -\frac{g \hat{E}}{\Omega} + \frac{1}{\Omega}
   (\partial_t -\Gamma_{13}) \frac{1}{\Omega}
   (\partial_t -\Gamma_{12}) \frac{g \hat{E}\Omega}
   {\Omega^2 + \Gamma_{12}\Gamma_{13}}  \label{eqs12atomic}  \\
\sigma_{13} &=& \frac{ig}{\Omega} ( \partial_t -\Gamma_{12})
  \frac{\hat{E} \Omega}{\Omega^2 + \Gamma_{12}\Gamma_{13}} \nonumber\\
&&  \hspace{0.5cm} + \frac{ig}{\Omega} (\partial_t -\Gamma_{12})
  \left[ \frac{\Gamma_{13}}{\Omega^2} \partial_t
  \frac{\hat{E} \Omega}{\Omega^2 + \Gamma_{12}\Gamma_{13}} \right. \nonumber \\
&& \hspace{1cm}  \left. + \frac{\Gamma_{12}}{\Omega} \partial_t
  \frac{\hat{E}}{\Omega^2 + \Gamma_{12}\Gamma_{13}} \right] \label{eqs13atomic}
\end{eqnarray}
These first order solutions are an excellent
approximation to the true solutions, as the
ratio between the probe field and the coupling
field is extremely small.
This can be seen by noting that within 
the context of quantum information storage the
probe pulse will contain only a few photons,
while, as we will see later, the coupling field
must generally be of the order of kW/cm$^2$
in order to overcome the inhomogeneous 
broadening.

The analysis above yields the relevant atomic
equations of motion for a single ion with
a specific detuning defined by its position
within its host. As we are dealing
with a collective effect, i.e. the incoming probe
pulse excites many ions at different sites simultaneously,
we need to average over the inhomogeneous broadening,
accounting for all possible detunings. Making the
assumption that the inhomogeneous broadening is
given by a Lorentzian, the averaged atomic
quantities are given by
\begin{equation}
\bar{\sigma}_{ij} = \frac{W_{12} W_{13}}{\pi^2 } \iint 
   \frac{d \Delta\omega_{12}}{(\Delta\omega_{12})^2 + W_{12}^2}
   \frac{d \Delta\omega_{13}}{(\Delta\omega_{13})^2 + W_{13}^2}
   \sigma_{ij}
\end{equation}
where the $\Delta\omega_{ij}$ are given by 
(\ref{eqdeltaomega13}) -- (\ref{eqdeltaomega12}), and 
$W_{12}, \,W_{13}$ are the inhomogeneous widths
of the $| 1\rangle \rightarrow |2\rangle$ and the
$| 1\rangle \rightarrow |3\rangle$ transitions
respectively.

The integration results in rather involved expressions,
and in order to make the analysis more tractable we make 
the following assumptions: Both the probe and coupling
fields are on resonance with an isolated ion (that is, 
$\Delta_{ij} = 0$), and $\gamma_{ij} \ll W_{ij}$
(requiring that the inhomogeneous linewidth is far wider than
the unbroadened linewidth). The first assumption 
can easily be met by choice of laser frequency, and
the second is obviously met since the inhomogeneous
linewidth is composed of many superimposed unbroadened
linewidths. These assumptions result in the averaged
atomic expressions
\begin{eqnarray}
\bar{\sigma}_{12} &=& -\frac{g\hat{E}\Omega}{\Omega^2 + W_{12}W_{13}}
   -\frac{1}{\Omega^2} \frac{\partial}{\partial t}
   \frac{g\hat{E}\Omega(-\gamma_{13}\Omega^2 +\gamma_{12} W_{13}^2)}
	{(\Omega^2 + W_{12}W_{13})^2} \nonumber \\
   && -\frac{1}{\Omega} \frac{\partial}{\partial t}
     \frac{g \hat{E}(-\gamma_{12}\Omega^2 +\gamma_{13} W_{12}^2)}
 	{(\Omega^2 + W_{12}W_{13})^2} \label{eqFullS12NAdef}\\
\bar{\sigma}_{13} &=&  \frac{-ig \hat{E}(-\gamma_{12}\Omega^2 +\gamma_{13} W_{12}^2)}
 	{(\Omega^2 + W_{12}W_{13})^2} 
	- \frac{ig}{\Omega^3} \frac{\partial}{\partial t}
 	\frac{\hat{E}\Omega W_{12} W_{13}}{\Omega^2 + W_{12}W_{13}} \nonumber \\
&& - \frac{ig}{\Omega^2} \frac{\partial}{\partial t}
 	\frac{\hat{E} W_{12}^2}{\Omega^2 + W_{12}W_{13}}
	+ \frac{ig}{\Omega} \frac{\partial}{\partial t} 
	\frac{\hat{E} \Omega}{\Omega^2 + W_{12}W_{13}} \nonumber \\
&&+ \frac{ig}{\Omega} \frac{\partial}{\partial t}
  \frac{1}{\Omega} \frac{\partial}{\partial t}
    \frac{\hat{E} (-\gamma_{12}\Omega^2 +\gamma_{13} W_{12}^2)}
    {(\Omega^2 + W_{12}W_{13})^2} \nonumber \\
&&  + \frac{ig}{\Omega} \frac{\partial}{\partial t}
  \frac{1}{\Omega^2} \frac{\partial}{\partial t}
    \frac{\hat{E} \Omega(-\gamma_{13}\Omega^2 +\gamma_{12} W_{13}^2)}
    {(\Omega^2 + W_{12}W_{13})^2}.
\end{eqnarray}

\section{Solutions}

We closely follow the analysis of Fleischhauer and Lukin
\cite{fleischhauerET2002}.

As a starting point we introduce the two quantum fields
$\hat{\Psi}$ and $\hat{\Phi}$. They are defined to be
\begin{eqnarray}
\hat{\Psi} &=& \cos \theta \hat{E} - \sin \theta \sqrt{N} \bar{\sigma}_{12} \label{eqpsidef2}\\
\hat{\Phi} &=& \sin \theta \hat{E} + \cos \theta \sqrt{N} \bar{\sigma}_{12}\label{eqphidef2}
\end{eqnarray}
with the inverse relations
\begin{eqnarray}
\hat{E} &=& \cos \theta \hat{\Psi} + \sin \theta \hat{\Phi} \label{eqEdef2}\\
 \sqrt{N} \bar{\sigma}_{12} &=& -\sin \theta \hat{\Psi} + \cos \theta \hat{\Phi}. \label{eqsigmadef2}
\end{eqnarray}
The time-dependent mixing angle $\theta(t)$ is defined by
\begin{equation}
\tan\theta = \frac{g \sqrt{N}}{\Omega(t)} \label{eqthetadef}
\end{equation}
Both $\hat{\Psi}$
and $\hat{\Phi}$ have bosonic commutation
relations in the limit of few photons and many atoms.
The action of $\hat{\Psi}^{\dagger}$ on the vacuum
creates dark states \cite{fleischhauerET2002}, which
contain no component of the excited state $|3\rangle$,
and are therefore unaffected by spontaneous emission
\cite{arimondo1996}. $\hat{\Phi}$, on the other hand,
creates states which couple to state
$|3\rangle$ and which are therefore lossy.
Consequently $\hat{\Psi}$ and 
$\hat{\Phi}$ are termed dark state 
and bright state polaritons respectively.

It is now clear that provided the system remains purely described
by the excitation $\hat{\Psi}$,
by rotating the mixing angle from $0$ to $\pi/2$
(equivalent to taking the control field from $\Omega=\infty$
to $\Omega = 0$) one can losslessly transfer 
the quantum probe field
into an atomic spin coherence, trapping the probe
field in the medium. Ramping the control field back
up rotates the spin coherence back into the probe
field which is then released and able to
propagate further.

Utilizing (\ref{eqEprop}) along with 
(\ref{eqFullS12NAdef})--(\ref{eqsigmadef2}),
one can obtain
\begin{eqnarray}
\left( \partial_t + c\cos^2 \theta \partial_z \right) \hat{\Psi}
&=&  -\dot{\theta} \hat{\Phi} - c\sin\theta\cos\theta
 	\partial_z \hat{\Phi} \nonumber \\
&& \hspace{-1.5cm} + g\sqrt{N}\sin\theta \left[
   \frac{(-\Omega^2 \gamma_{12} + W_{12}^2 \gamma_{13})\hat{E}\Omega}
    {(\Omega^2 + W_{12}W_{13})^2} \right. \nonumber\\
&& \hspace{-1.5cm} + \frac{W_{12} W_{13}}{\Omega^2}
	 \frac{\partial}{\partial t}
    \frac{\hat{E}\Omega}{\Omega^2 + W_{12}W_{13}} \nonumber \\
&&  \hspace{-1.5cm} + \left. \frac{W_{12}^2}{\Omega} 
	\frac{\partial}{\partial t}
    	\frac{\hat{E}}{\Omega^2 + W_{12}W_{13}} \right] 
		\label{eqWithFullNAcorrections}
\end{eqnarray}

We make the assumption that $W_{12} \ll W_{13}$, 
that is, that the 
inhomogeneous width of the ground state is much
less than the width of the upper, excited,
state. In solids the upper state broadening
is normally several orders of magnitude larger
than that of the ground state, making this an
excellent approximation. We can thus neglect
the the last term in (\ref{eqWithFullNAcorrections})
relative to the second to last term.

After using (\ref{eqEdef2}) to eliminate $\hat{E}$ and carrying 
out the differentiation, Eq.
(\ref{eqWithFullNAcorrections}) can be written
\begin{eqnarray}
\left( \partial_t + c\cos^2 \theta \partial_z \right) \hat{\Psi} &=& 
-\dot{\theta} \hat{\Phi} - c\sin\theta\cos\theta \partial_z \hat{\Phi} \nonumber\\
&& \hspace{-2.7cm} + \tan \theta (\cos\theta\hat{\Psi} 
		+ \sin \theta \hat{\Phi})
	\left[ \frac{\Omega^2 \sin \theta(W_{12}^2 \gamma_{13}
	-\Omega^2 \gamma_{12})}{(\Omega^2 + W_{12}W_{13})^2} \right. \nonumber \\
&&	\left. \hspace{-2.3cm} + \frac{\dot{\theta}}{\cos\theta} 
	\frac{W_{12} W_{13}(\Omega^2 - W_{12}W_{13})}
	{(\Omega^2 + W_{12}W_{13})^2} \right] \nonumber \\
&& 	\hspace{-2.6cm} + \frac{\sin^2 \theta W_{12} W_{13}}{\Omega^2 + W_{12}W_{13}}
	\left( \dot{\hat{\Psi}} + \tan\theta \dot{\hat{\Phi}}
	- \tan\theta \dot{\theta} \hat{\Psi} 
	+ \dot{\theta} \hat{\Phi} \right) \label{eqphiEOM}
\end{eqnarray}
where we have used both $\Omega$ and $\theta$ even though
they are related, as this makes the expression more compact.
The first two terms on the right hand side are identical 
to those present in the gaseous medium considered in 
Ref. \cite{fleischhauerET2002}, where there is no
inhomogeneous broadening and the ground state
coherence lifetime $\gamma_{12}$ is taken to be
infinitely long. The remainder of the terms, however,
are distinct to the case of a solid medium.

To obtain the final equation of motion for $\hat{\Psi}$,
we need to eliminate $\hat{\Phi}$ from (\ref{eqphiEOM}).
This can be accomplished by using (\ref{eqFullS12NAdef})
and (\ref{eqphidef2}) to obtain
\begin{eqnarray}
\hat{\Phi} &=& \frac{g\sqrt{N}}{\Omega} \cos\theta \left[ 
 \frac{\hat{E} W_{12}W_{13}}{\Omega^2 + W_{12}W_{13}} \right. \nonumber \\
&& -\frac{1}{\Omega} \frac{\partial}{\partial t}
   \frac{\hat{E}\Omega(-\gamma_{13}\Omega^2 +\gamma_{12} W_{13}^2)}
	{(\Omega^2 + W_{12}W_{13})^2} \nonumber \\
&& \left.   \hspace{1cm} -\frac{\partial}{\partial t}
   \frac{\hat{E}(-\gamma_{12}\Omega^2 +\gamma_{13} W_{12}^2)}
 	{(\Omega^2 + W_{12}W_{13})^2} \right].
	\label{eqPhidef}
\end{eqnarray}
Again making the replacement
$\hat{E} = \cos \theta \hat{\Psi} + \sin \theta \hat{\Phi}$
and performing the derivatives one finds the relation
\begin{eqnarray}
\hat{\Phi} &=& \left[ \frac{W_{12} W_{13} \sin \theta \cos\theta}
	{\Omega^2 + W_{12}W_{13}} 
	 - (\alpha + \beta) \dot{\theta} \right] \hat{\Psi} 
		+ \beta \cot \theta \dot{\hat{\Psi}} \nonumber \\
&& \hspace{.5cm}+ \left[ \frac{W_{12} W_{13} \sin^2 \theta}
	{\Omega^2 + W_{12}W_{13}} - \alpha\dot{\theta} 
		\tan\theta \right] \hat{\Phi} \label{eqPhi}
\end{eqnarray}
where
\begin{eqnarray}
\alpha &=& \frac{\gamma_{13} \Omega^2 (3 W_{12}W_{13} - \Omega^2) + \gamma_{12} W_{13}^2(3\Omega^2 - W_{12} W_{13})}{(\Omega^2 + W_{12}W_{13})^3} \nonumber \\
\beta &=& \frac{\sin^2 \theta((\gamma_{12} + \gamma_{13})\Omega^2 - 
	\gamma_{12} W_{13}^2 - \gamma_{13} W_{12}^2)}
	{(\Omega^2 + W_{12}W_{13})^2}
\end{eqnarray}
From Eq. (\ref{eqPhi}) it can be shown that to keep from populating
the bright state polariton, that is to ensure that $\hat{\Phi}$
is small relative to $\hat{\Psi}$, we require that
\begin{equation}
\Omega^2 \gtrsim 3 W_{12} W_{13}. \label{eqPowerCondition}
\end{equation}
Thus to keep the bright state from being populated the
strength of the control field $\Omega$ must always 
dominate the inhomogeneous broadening.

If this criterion is met, we have
\begin{equation}
\hat{\Phi} = \left[ \frac{W_{12} W_{13} \sin \theta \cos\theta}
	{\Omega^2 + W_{12}W_{13}} 
	 - (\alpha + \beta) \dot{\theta} \right] \hat{\Psi} 
		+ \beta \cot \theta \dot{\hat{\Psi}}
\label{eqPhiApprox}
\end{equation}

The dynamics of the dark state polariton divide naturally
into two subcases --- one where the control field is altered
so slowly that the evolution of the atomic states exactly
follows the control field, and a second where some element
of nonadiabaticity is considered.

\subsection{Adiabatic case}

In totally adiabatic evolution, only the first term
of (\ref{eqPhiApprox}) is relevant. Thus
\begin{equation}
\hat{\Phi}_{\textrm{ad.lim.}} = \frac{
    g\sqrt{N} \Omega W_{12}W_{13} }{(\Omega^2 + g^2 N)
	(\Omega^2 + W_{12} W_{13})} \hat{\Psi} \label{eqPhiAdLim}
\end{equation}
Provided one remains in the regime given by 
(\ref{eqPowerCondition}) it is clear that
$\hat{\Phi}$ can be neglected relative to $\hat{\Psi}$.

We note that Eq. (\ref{eqPhiAdLim}) should 
contain a Langevin (vacuum)
noise term so that the commutation relations are met.
However since $\langle \hat{\Phi}^{\dagger} \hat{\Phi}\rangle \ll
 \langle \hat{\Psi}^{\dagger} \hat{\Psi} \rangle$,  
the bright state polariton is never appreciably
excited. It is thus possible to adiabatically transfer
the electromagnetic probe pulse into an atomic dark
state with no component projected onto the upper excited
state, therefore avoiding destruction and noise due
to spontaneous emission.

We are now able to derive the equation of motion for 
the dark state polariton in the adiabatic limit
by using (\ref{eqphiEOM}), ignoring the $\hat{\Phi}$ terms
and noting that as we are changing the control
field adiabatically $\dot{\theta} = 0$. This yields
\begin{equation}
\left( \partial_t + c\cos^2 \theta \partial_z \right) \hat{\Psi} =
	\frac{W_{12} W_{13} \sin^2 \theta}{\Omega^2 + W_{12}W_{13}} \dot{\hat{\Psi}}
	 - \sin^2\theta \Gamma_{\Psi} \hat{\Psi}
	\label{eqPsiAdLimEOM}
\end{equation}
with
\begin{equation}
\Gamma_{\Psi} = \frac{\Omega^2 (\Omega^2 \gamma_{12}
		 - W_{12}^2 \gamma_{13})}{(\Omega^2 + W_{12}W_{13})^2}.
		\label{eqGammaPsiDef}
\end{equation}
This result should be compared with that 
obtained using a gaseous medium, where no
inhomogeneous broadening is present, and 
the ground state coherence time is taken to 
be infinitely long \cite{fleischhauerET2002}:
\begin{equation}
\left( \partial_t + c\cos^2 \theta \partial_z \right) \hat{\Psi} = 0	 
	\label{eqPsiAdLimEOMGasCase}
\end{equation}

The first term on the right hand side of 
(\ref{eqPsiAdLimEOM}) is
clearly a correction to the group velocity
of the polariton pulse. Provided we remain
in the regime given by (\ref{eqPowerCondition})
this results in a velocity correction factor
close to unity.

It is equally clear that $\Gamma_{\Psi}>0$ and so
denotes a loss term. Furthermore, within the
regime (\ref{eqPowerCondition}), $\Gamma_{\Psi}$
is bounded by $\gamma_{12}$, the dephasing
rate of the ground state coherence. This
is logical, and indicates that the maximum
storage time is limited by the lifetime
of the ground state coherence.

One difficulty remains: Because the power of
the control field must dominate the inhomogeneous
broadening, we cannot reduce it to zero in 
order to achieve a zero group velocity and
stop the probe pulse. The minimum velocity
occurs when $\Omega^2 \approx W_{12} W_{13}$
and is given by
\begin{equation}
v_g = c \cos^2 \theta_{\mathrm{min}} = \frac{c W_{12} W_{13}}{W_{12} W_{13} + g^2 N}.
\end{equation}
Thus, in order to achieve a near-zero group
velocity, one requires $g^2 N \gg W_{12} W_{13}$.

In general, the solids of interest consist of rare-earth ion
doped crystals \cite{kuznetsovaET2002}. Consequently, 
$W_{12}$ is of the order of
tens of kilohertz, and $W_{13}$ is of the order of gigahertz.
$g^2 N$, on the other hand, tends to be within a few orders
of magnitude of $\sim 10^{21}\,$Hz$^2$. (These assumptions,
along with solids other than rare-earth ions doped into 
crystal hosts, will be considered in greater detail 
in Section \ref{secDiscussion}.) The minimum polariton
velocity is thus perhaps few tens of meters per second,
although this is highly dependant on the medium chosen.
Similarly, at the minimum control field strength
$\sin\theta \approx 1-W_{12}W_{13}/g^2 N$, 
indicating that practically all of the probe
field has been transferred and stored.

This conclusion reproduces the gaseous medium
result: a few-photon input pulse
can have its quantum state stored as a spin
coherence, provided the control field changes
sufficiently slowly, and with a storage time
bounded by the decay time of the ground state
coherence. The group velocity of the polariton
is $c\cos^2 \theta$, and approaches zero as the
control field is reduced, ensuring that the
pulse is slow enough to be considered stored. 

To achieve this, we required two additional
conditions that are a consequence of working
in a solid:
\begin{eqnarray}
g^2 N & \gg &  W_{12} W_{13}\label{eqggtW} \\
\Omega^2 & \gtrsim & 3 W_{12}W_{13} \label{eqWcond}.
\end{eqnarray}
To what extent these conditions can be met in
current materials will be considered in Section
\ref{secDiscussion}.


\subsection{Nonadiabatic corrections}
\label{subsecNonadiabaticCorrections}

Although in principle one can modify the control
field as slowly as one desires, and thus ensure
that one remains in the adiabatic regime, this is
not realistic for practical quantum information 
storage. As the storage time is bounded by the
ground state coherence lifetime, one must at
a minimum be able to complete a storage and
retrieval operation within this period. This puts
a lower bound on how slowly the control field
can be turned off and on. Consequently one must
determine the maximum speed at which the control
field can be reduced to zero without
nonadiabatic effects destroying the storage process.
It is known that these nonadiabatic losses
can be made neglible in a gaseous medium;
we now consider whether the same can be made
to hold in a solid medium.

As we wish to include first order nonadiabatic
corrections, we cannot ignore all time derivatives
as we did in the previous section.
Making use of (\ref{eqPhiApprox}) and (\ref{eqphiEOM})
we obtain the following equation of motion
equation of motion for the dark state polariton:
\begin{equation}
\left( \partial_t + c\cos^2 \theta \partial_z \right) \Psi =
  -A(t) \Psi + B(t) c \frac{\partial}{\partial z} \Psi
   + C(t) c^2 \frac{\partial^2}{\partial z^2} \Psi
	\label{eqEOMInTermsOfABC}
\end{equation}
where
\begin{eqnarray}
A(t) &=& (1+\gamma) \sin^2\theta \Gamma_{\Psi} 
	+ \dot{\theta}
	\left( \gamma\cot\theta + \gamma \tan \theta \nonumber \right. \\ 
&& \hspace{.1cm} \left. - (\alpha+\beta) \tan\theta \sin^2\theta
	 \Gamma_{\Psi} - (1+\gamma)\delta \tan \theta
	 \right. \nonumber \\
&& \hspace{.3cm} \left. - 2 \gamma^2 \cot \theta  + \gamma^2 \tan \theta 
	- 2g^2 N \gamma^2 \cot\theta\csc^2\theta) \right) \nonumber \\
&&  \hspace{0.8cm} - \dot{\theta}^2 (\alpha+\beta)
	(1-\gamma-\delta\tan^2 \theta) \label{eqAdef} \\ 	
B(t) &=& -2 \gamma \cos^2 \theta - \beta \Gamma_{\Psi} 
	\sin^2 \theta \cos^2 \theta 
	+ \dot{\theta} \sin\theta\cos\theta \left( \alpha \right. \nonumber \\
&& \left. + \beta (1 + \cot^2 \theta - \delta - 
	\gamma \cot^2\theta)\right) \label{eqBdef} \\
C(t) &=& \beta \cos^4 \theta \label{eqCdef}
\end{eqnarray}
and where 
\begin{eqnarray}
\gamma &=& \frac{\sin^2 \theta W_{12} W_{13}}{\Omega^2 + W_{12} W_{13}} \\
\delta &=& \frac{W_{12} W_{13}(\Omega^2 -W_{12} W_{13})}
	{(\Omega^2 + W_{12} W_{13})^2}.
\end{eqnarray}
$A(t)$ and $C(t)$ represent losses, and $B(t)$
represents a modification of the group velocity
of the polariton. 

To determine whether the transfer of the probe pulse
into a trapped dark state can occur within an 
interval significantly shorter than the storage time,
we must calculate how large $\dot{\theta}$ can be
without incurring significant nonadiabatic losses.
We will follow the analysis of Ref. \cite{fleischhauerET2002}. 

Eq. (\ref{eqEOMInTermsOfABC}) can be solved by
making the Fourier transform
$\Psi(z,t) = \int dk \tilde{\Psi}(k,t) e^{-ikz}$. This gives
\begin{eqnarray}
\tilde{\Psi}(k,t) &=& \tilde{\Psi}(k,0) \exp \left[ ik\int_0^t dt'
	(v_{gr}(t') -cB(t'))\right] \nonumber \\
&& \times \exp \left[ \int_0^t dt'
	(A(t') - k^2 c^2 C(t'))\right]
\end{eqnarray}
where the first term is a group velocity
correction and the last term contains the
nonadiabatic losses and pulse-spreading effects
we are interested in. To avoid losses, 
the integral in the exponent
must be small relative to one. This results in
the two conditions
\begin{eqnarray}
\int_0^{\infty} dt' A(t') &\ll& 1 \label{eqCondOnA} \\
k^2 c^2 \int_0^{\infty} dt' C(t') &\ll& 1. \label{eqCondOnc}
\end{eqnarray}
Since $\Omega^2 > 3 W_{12}W_{13}$ we can construct an
upper bound for $C(t)$ which gives
\begin{equation}
k^2 c^2 \gamma_{13} \int_0^{\infty} dt' \frac{\sin^4 \theta \cos^2\theta}{g^2 N} \ll 1.
\end{equation}
This is identical to the
condition derived for a gaseous medium, and can
be shown to be equivalent to the condition
that \cite{fleischhauerET2002}
\begin{equation}
z \ll \frac{g^2 N }{\gamma_{13} L_p^2}
\label{eqInitialBandwidth}
\end{equation}
where $L_p$ is the length of the probe pulse
in the medium (i.e. after compression due to 
EIT effects) and $z$ is the distance the pulse
travels in the medium before being completely
stored, and can be seen as a lower bound on 
the medium length required. This condition
in turn is equivalent to requiring only that
the initial spectral width of the pulse 
before beginning deceleration fits within
the initial EIT transparency window.

We turn now to the condition given by 
(\ref{eqCondOnA}). Looking at
(\ref{eqAdef}) the first term merely states
that the polariton cannot be stored longer
than the lifetime of the ground state 
coherence $1/\gamma_{12}$.

Next we consider the term proportional to
$\dot{\theta}^2$. We take the initial control
field strength to be $\Omega_0$ and parameterize
the final control field strength $\Omega(\tau)$ 
as $k= \Omega(\tau)/\sqrt{W_{12} W_{13}}$. 
Assuming a linear decrease in $\Omega$ 
over the time $\tau$, then integration of
the term proportional to $\dot{\theta}^2$
with respect to time yields the condition
\begin{equation}
\tau \gg \frac{\gamma_{13} \Omega_0}{k^7(W_{12} W_{13})^{3/2}}
\label{eqAnotherNAcond}
\end{equation}
for $1 < k \lesssim 10$. This puts an upper
bound on the speed one can reduce the 
control field field.

Finally we consider the term in (\ref{eqCondOnA})
proportional to $\dot{\theta}$. As the integral is
taken with respect to time, the presence of the
$\dot{\theta}$ term ensures that there is no
time dependence in the result. Thus effectively the
integral is carried out with respect to $\theta$,
and results in an overall loss factor. This
loss factor is relatively insensitive to the
precise values of all the parameters excepting
the final control field strength $\Omega(\tau)$.
Again if $k = \Omega(\tau)/\sqrt{W_{12} W_{13}}$
then the loss factor is approximately
\begin{equation}
\eta = \exp \left[ \frac{3+2k^2}{(1+k^2)^2} + 2\ln \frac{k^2}{1+k^2}\right].
\label{eqSupressionFactor}
\end{equation}

These are the conditions that must be met if we are to stop
and store a probe pulse within a solid. Whether
they can be met is strongly dependent on the
solid that is used, and it is to this that we
now turn.

\section{Practical considerations and examples}
\label{secDiscussion}

The solids most often considered within the context 
of coherent optical behavior are rare-earth doped crystals.
For a review of the general properties of these systems we
refer to the recent paper by MacFarlane \cite{macfarlane}.
The dopant rare-earth ions are characterized by 
low inhomogeneous broadening of their lower state
hyperfine transitions, well-characterized energy
levels, and existence of a zero-phonon line at
low temperatures that coincides with the
commonly used $f$-$d$ transitions 
\cite{kuznetsovaET2002} with a reasonably large
transition dipole moment but still a relatively
narrow homogeneous optical width.

In addition,
they are attractive for quantum information storage
due to their high ratio of optical-transition
inhomogeneous broadening to spin-transition
inhomogeneous broadening, which allows
the writing of many discrete channels via
spectral hole burning \cite{mossberg}
and pulse compression by photon echo effects
\cite{wang}.  

As has been made clear from the foregoing analysis,
there are two primary quantities that govern the
ability to transfer the quantum information from
the probe pulse into a spin coherence. They are
$W_{12}W_{13}$, the product of the inhomogeneous
widths of the optical and spin transitions, and
$g^2 N$, the collective coupling strength of the
medium.

The single-atom coupling is given by 
$g = d_{13}\sqrt{\nu/2\hbar \epsilon_0 V}$.
The dipole moments for rare earths generally
lie in the range $10^{-29}$ -- $10^{-32}$ Cm. 
We choose $10^{-30}\,$Cm as a representative
value in the following. Assuming a wavelength
of $1000\,$nm, we find the convenient relation
that 
\begin{equation}
g^2 N \,[{\mathrm{Hz}}^2] \sim \frac{N}{V} \, [{\mathrm{m}}^{-3}],
\end{equation}
that is, the collective coupling strength
is simply given by the density of the dopant
atoms in the medium. The density of
rare-earths dopant ions in crystals can easily be as
high as $10^{17}$ -- $10^{19}\,$cm$^{-3}$,
depending on the dopant and matrix material.
Thus $g^2 N \sim 10^{23}$ -- $10^{25}\,$Hz$^2$,
many orders of magnitude higher than what is
possible in gases.

The magnitude of the optical and spin inhomogeneous
broadening is strongly dependent on the rare-earth
and on the electronic transition chosen. A
typical range of values for $W_{13}$ is 
1--10 GHz, while $W_{12}$ ranges from 100 Hz -- 1 Mhz. 
Consequently, one could expect 
$W_{12} W_{13} \sim 10^{15}\,$Hz$^2$,
and it is therefore clear that the condition 
$g^2 N \gg W_{12} W_{13}$ is very easily met
in these materials.

We now consider the power requirements of the
coupling laser. If we wish to let the probe
pulse enter the medium at speed $c$, and 
then reduce the coupling field strength to
its minimum value, effectively stopping the
pulse, it is necessary to rotate the mixing
angle $\theta$ from 0 to $\pi /2$. A value of
$\theta=0$ corresponds to a coupling field
of infinite intensity, or more realistically,
$\Omega^2 \gg g^2 N$. Optimistically assuming 
$W_{12} W_{13} = 10^{15}\,$Hz$^2$ and $g^2 N = 10^{17}\,$Hz$^2$,
we might require $\Omega^2(0) = 10^{19}\,$Hz$^2$. Using
\begin{equation}
I = \frac{\Omega^2 \hbar^2 c \epsilon_0}{2 d_{13}^2}
\end{equation}
and assuming a dipole moment of $d_{13} = 10^{-30}\,$Cm 
we see that this corresponds to a coupling laser
intensity of 10 kW/cm$^2$. These numbers are only
indicative, and it is possible
to reduce the power requirements by choosing 
systems with smaller inhomogeneous broadening
and reducing the dopant concentration. 

One must also take into consideration
the length of medium required to stop the pulse.
Naively, if the coupling laser intensity is reduced from 
$\Omega(0) \gg g\sqrt{N}$ to its
minimum value $\Omega(\tau)\sim \sqrt{W_{12} W_{13}}$
in time $\tau$, the distance the
pulse travels is
\begin{equation}
z=\int_0^\tau c \cos^2\theta(t) dt \approx c\tau.
\end{equation}
Thus, bearing in mind the adiabaticity requirements,
if $\tau \sim 10^{-6}\,$s, the stopping distance
is 300m. This may just be feasible for a experiment
with a doped fiber, but certainly not for a crystal.

The correct approach, which obviates this
difficulty as well as reducing the pump power
required, is to
ensure the coupling field has a strength such that 
the probe pulse is in the slow group velocity
regime as soon as it enters the medium, namely
$W_{12} W_{13} \ll \Omega(0)^2 \ll g^2 N$. In
this case
\begin{equation}
z=\frac{\Omega(0)^2}{3 g^2 N} c\tau.
\end{equation}
As it is possible to make $g^2 N$ extremely
large in a solid, there is no difficulty
in stopping the probe pulse within a few
centimeters. The initial coupling laser
Rabi frequency can now be orders of magnitude
lower, provided it still dominates the 
inhomogeneous broadening, resulting in
an initial coupling laser intensity of 
$\sim 100$ W/cm$^{2}$.

The final conditions that must be met are
the adiabaticity criteria, namely
(\ref{eqInitialBandwidth}),
(\ref{eqAnotherNAcond}) and 
(\ref{eqSupressionFactor}), which
exist only when one moves away from the
adiabatic limit.

The first condition is that probe pulse bandwidth
after entry to the medium must be within the EIT
transparency window before the coupling laser
intensity is reduced. In the case of a strong
coupling field in a solid, the EIT window is
given by $\Gamma_{EIT} = \Omega^2 / W_{13}$
\cite{kuznetsovaET2002}, corresponding to a
bandwidth of $\sim 10^6$ -- $10^8\,$Hz, 
depending on the intial coupling field strength.
If a broader bandwidth is required, one merely
needs to increase the coupling field strength.

Eq. (\ref{eqAnotherNAcond}) is a fairly weak
condition. Using the numbers
$W_{12} W_{13} \sim  10^{15}\,$Hz$^2$,
$\Omega(0)^2 \sim 10^{17}\,$Hz$^2$ and 
$\gamma_{13} \sim 10^7\,$Hz one obtains
$\tau\gg 10^{-7}$s, which is certainly still orders
of magnitude shorter than the limit of the 
storage time which is governed by $1/\gamma_{12}$. 

The final condition (\ref{eqSupressionFactor})
is highly dependent on to what extent the
final control field strength dominates the
inhomogeneous broadening. Taking
$\Omega(\tau) = 3 \sqrt{W_{12} W_{13}}$ gives
a damping factor of $\eta \approx \exp[-0.0007]$,
which is negligible.

Thus, in broad, it appears that quantum
information storage using this technique
is feasible in rare-earth doped crystals.
It is not clear, however, that there is
one type of material that possesses 
all the properties that would make it an
ideal candidate. The oscillator strength
of the rare-earth itself is not too important, 
as it can generally be compensated for by altering
the dopant density. The crucial quantities
are the inhomogeneous broadening widths
$W_{12}$ and $W_{13}$, the collective
coupling $g^2 N$ and the dephasing rate 
$\gamma_{12}$. The ideal material would
have a low $W_{12} W_{13}$, a high $g^2 N$,
and a very low $\gamma_{12}$.

Some measures can be taken to reduce
$W_{13}$. For example, Ham {\textit{et al.}}
introduce a repump laser to prevent
spectral hole-burning by the coupling and
probe lasers, and consequently limit the
optical inhomogeneous broadening to the
repump laser jitter ($\sim 1$ MHz) \cite{hamET1997}.
In the scheme described in this paper, we have 
assumed very weak probe fields
for quantum information purposes, and so only a
tiny fraction of the atoms make the transition
to state $|2\rangle$, rendering such a repump
laser unnecessary. Similar spectral
hole-burning techniques, however, could be used prior
to applying the probe pulse, 
selecting a subset of the ions within a 
particular spectral range  \cite{macfarlane}
and thus drastically
reducing $W_{13}$. The drawback is a reduction
in the interacting ion density, but as 
$g^2 N \sim 10^{23}\,$Hz$^2$ is attainable, reducing
the density by a factor of 1000 is certainly
acceptable for a similar 1000-fold reduction in the
inhomogenous broadening.

The storage time, that is the time one may
wait before the quantum field is released
by the turning on of the strong pump field,
is limited both by the homogeneous width 
$\gamma_{12}$ and the inhomogeneous width 
$W_{12}$ of the lower hyperfine transition 
in the ions. $\gamma_{12}$ serves as an
absolute lower limit as discussed in the
previous section. $W_{12}$ is a limit due
to the fact that the phases of different ions
evolve at different speeds due to inhomogeneity,
meaning that after a time $1/W_{12}$ the
stored information will no longer be
coherent. In principle this can be overcome.
Spin echo techniques have been exploited to
compensate inhomogenous frequency shifts and to
observe features limited only by the homogeneous 
lower state width in ion doped crystals \cite{wang}. 
In our case, however, it is not plausible that 
one can precisely invert the populations to
a level of precision matching the almost
insignificant number of photons stored in the
medium. Thus in practice one is limited by
the storage time $1/W_{12}$ rather than
the longer storage time $1/\gamma_{12}$.
It has been shown, however, that strong magnetic
bias fields can reduce the inhomogeneous
broadening significantly, and suggests that
storage times of the order of 100 ms or more
may be achievable \cite{sellars}. 

For integration with current telecommunication 
technologies, it is natural to speculate about
the possibilities to slow and store light in
doped optical fibres and wave guides rather than
crystals. In fibres and wave guides, where the
ions are doped into a glass host, the inhomogeneous
widths of both the optical and the hyperfine
transitions are much larger than in crystals
\cite{cormierET1993,desurvireET1990,szeftelET1975}. Persistent hole burning
has been demonstrated in glass fibres \cite{yano}, and a natural strategy 
thus seems to be a preparation of the system by pumping of all ions in a broad
frequency range to passive spectator levels, 
leaving only ions which have their $1-3$ and $1-2$ transitions in desirable
frequency windows in the middle of this range in their state
$|1\rangle$. 
Considerable improvement of the hole burning must be achieved
and further understanding of the homogeneous
width of the transitions is clearly needed before serious attempts
along this line can be carried out.
As commented upon above, hole burning leads to a significant
reduction of the number of ions available for the light storage.
A crystal fibre may be doped only in the central rod
which  forms the central wave guide in the fibre \cite{russel96}. 
The light may thus be confined to a cross section about the size of
the (resonant) absorption cross section of a single ion which, 
together with the achievable lengths of these fibres, may compensate for
the low concentration, and make slowing and storage of light possible.
It would also be possible to set up an optical cavity by writing 
a Bragg grating in the fibre \cite{kristensen} (or by coating the
faces of the crystals in our main proposal) and in this way enhance 
the interaction of
the field with the atomic system, as it has been proposed for 
free atoms and for ions \cite{fyl}. An analysis of this proposal
lies beyond the purpose of the present paper.
Significant non-linear dynamics, for example super-continuum generation,
has been observed in crystal fibres at very high light intensities as
a consequence of the non-linear susceptibility of the glass host
\cite{russel01}.
For the above analysis to hold, one should avoid this parameter regime,
but we do not exclude the possibility that useful effects may be derived
from the optical nonlinearity of the host material in conjunction
with the EIT dynamics due to the dopant ions.

\section*{Acknowledgements}
We would like to thank Michael Fleischhauer 
for his helpful suggestions.

\def\etal{\textit{et al.}}

\vspace{.5cm}

\end{document}